%% file: editor.tex
\documentclass[graybox, envcountchap]{svmult}

\usepackage{makeidx}         
\usepackage{graphicx}        
\usepackage{multicol}        
\usepackage[bottom]{footmisc}

\usepackage{newtxtext}       %
\usepackage{newtxmath}       

\makeindex             

\usepackage{import}

\begin{document}




\mainmatter

\import{author1/}{author}
%

\backmatter
\appendix
\include{appendix}
\include{glossary}
\printindex


\end{document}

%% file: author1/author.tex
%
%
%

%
%
%
%
%
%
%

\title{Does publicity in the science press drive citations? A vindication of peer review}
\author{Manolis Antonoyiannakis}
\institute{Manolis Antonoyiannakis$^{1,2}$ \at $^1$Department of Applied Physics \& Applied Mathematics, Columbia University, 500 W. 120th St., Mudd 200, New York, NY 10027, \email{ma2529@columbia.edu}
\at $^2$American Physical Society, Editorial Office, 1 Research Road, Ridge, NY 11961-2701 
}
%
%
\maketitle

\abstract{We study how publicity, in the form of highlighting in the science press, affects the citations of research papers. After a brief review of prior work, we analyze papers published in {\it Physical Review Letters} (PRL) that are highlighted across eight different platforms. Using multiple linear regression we identify how each platform contributes to citations. 
We also analyze how frequently the highlighted papers end up in the top 1\% cited papers in their field. We find that the strongest predictors of medium-term citation impact---up to 7 years post-publication---are Viewpoints in {\it Physics}, followed by Research Highlights in {\it Nature}, Editors' Suggestions in PRL, and Research Highlights in {\it Nature Physics}.
Our key conclusions are that (a) highlighting for importance identifies a citation advantage, which (b) is stratified according to the degree of vetting during peer review (internal and external to the journal).
This implies that we can view highlighting platforms as predictors of citation accrual, with varying degrees of strength that mirror each platform's vetting level.}

\section{Introduction}
\label{sec:1}

\noindent

\noindent
Publishing is a selection process. Through a series of selections, or filtering, journal editors decide which papers merit external review, which papers to send back to referees for a second (or third) look, and finally, which papers to publish. 

But increasingly over the past 20 years, the selection of papers does not stop at publication, and editors curate lists of their favorite accepted papers for their readers.
Nor is such filtering limited to a journal's own papers, as some publishers (notably the {\it Science} and {\it Nature} series journals) curate lists of papers published among other journals in their field.  So, why do editors produce these selections?
In view of an exponential increase in scientific publications~\cite{century-of-physics} and the wide accessibility of papers through online repositories, the role of journals and editors is being redefined. By curating lists of accepted or recently published papers, editors add value to their journals, mitigate diminishing attention spans and retain a larger portion of their readers' attention. Indeed, these select sets of papers, or {\it highlights}, are deemed to be of higher quality, interest, or importance than the typical paper in their source journals. Some publishers invest further editorial resources to increase visibility of these select papers and accompany them with short summaries by editors, or longer commentaries by experts. Depending on publisher, journal, and type, highlights are called Editors' Choice, Research Highlights, Editors' Suggestions, Perspectives, News \& Views, Viewpoints, etc. See Table 1. The community seems to pay attention. Indeed, researchers often promote (in their websites, resum{\'e}s, cover letters, and progress reports) their highlighted papers, while funding agencies track their grantees' progress by monitoring coverage in various highlighting platforms.

Highlighting is a form of publicity in the science press, aimed at an expert audience of peers, from research scientists to science journalists. It is a different kind of publicity than University press releases, which are promotions by interested parties, or coverage of research in popular media (newspapers, magazines), whose readership is different from that of scholarly journals. 

\noindent

Given the efforts publishers undergo to select and promote their favorite papers, one would expect that highlighted papers stand out compared to other papers in their source journals. In particular, when highlighting is done for importance (however problematic it is to define ``importance'') we would expect it to translate into higher citation counts. 
Whether the highlighting process identifies or {\it causes} increased citations is a pertinent question. 
Here, we find both effects are at play but have different magnitudes: Papers that are highlighted randomly pick up a few additional citations, but nowhere as many as those that are deliberately highlighted by editors for importance.
In contrast, when highlighting is done for an intrinsically interesting, cute or elegant paper, then citations may not be an appropriate measure of distinction. Citation metrics have been found to agree well with peer review at the aggregate level. For example, citation-based rankings of British universities correlate with the Research Excellence Framework (REF) power ranking at 0.97~\cite{REF-rainy-Sunday}, while another analysis reported correlations above 0.8 for a number of fields, including physics~\cite{metrics-peer-review-Traag}. 
Therefore, it makes sense to explore whether the additional 
``round'' of review on highlighting (beyond mere acceptance of a paper) correlates with citation counts of those select papers.

\noindent

In this chapter, we address these issues quantitatively. As a case study, we analyze citation data for several publicity markers on papers published in the journal {\it Physical Review Letters} (PRL) of the American Physical Society (APS). 
Building on our previous work~\cite{PRB-Editorial, highlighting-MM-2016, highlighting-MM-2017}, we analyze here a much broader set of 8 highlighting markers, both internal and external to the APS. We analyze the citation behavior of various markers of publicity, from journal cover to highlighting in the source journal to highlighting in non-APS journals such as {\it Nature} or {\it Nature Physics}. Using multiple linear regression, we explore whether publicity in the science press is associated with higher citations, so we can use it as an indicator of future citation impact. We also explore how frequently the highlighted papers end up in the highly-cited-papers list of the Essential Science Indicators in Clarivate Analytics~\cite{Clarivate-HC}, i.e., the top 1\% cited papers in their field, and whether we can use any of the highlighting markers as predictors of placement in this list. 


We make a distinction of whether the paper is highlighted by a platform that belongs to its own publisher (its source journal, a sibling journal, or a blog) or to another publisher. In the former case, we speak of an {\it intra-highlight}, in the latter of an {\it inter-highlight}. Such a distinction is significant because when the editors select one of their own papers (intra-highlight) they have access to more information from the paper's peer review (referee reports, discussions with the editorial board, caveats that arose in the review process, etc.) to assist their decision. In contrast, when the editors select a paper in another journal (inter-highlight), they generally have no detailed knowledge of the paper's peer review other than its conclusion, namely, that the paper was accepted in its source journal and, for a published paper, that it was highlighted there.
(Inter-highlighting editors have even less information if they select a paper from the arXiv for highlighting prior to acceptance or publication in any journal.)
Why do we bother to distinguish between intra- and inter-highlights? Because even among papers in the same journal there is typically considerable variation of quality, importance, or interest. These attributes come up usually in the review process, enriching the editorial decision making process and aiding the editors in selecting their journal's highlights---an advantage that is lost to inter-highlighting platforms.

\section{Previous work}
\label{sec:Previous}

As early as 1991, Phillips et al.~\cite{Phillips} reported a citation advantage for research articles in the {\it New England Journal of Medicine} that received press coverage in the {\it New York Times}. The effect lasted in each of the 10 years since publication, and was strongest in the first year. They also sought to determine whether publicity (coverage by the {\it Times}) itself increased citations, or merely earmarked outstanding articles that would have been well-cited anyway ({\it publicity} vs. {\it earmark} hypothesis). Due to a 12-week strike in 1978, the {\it Times} continued to print a limited edition but did not sell copies to the public. Scientific articles covered by the {\it Times} during the strike period were not cited more than the control group, thus giving support to the publicity hypothesis over the earmark hypothesis.

Kurtz et al.~\cite{Kurtz} analyzed data from the NASA Astrophysics Data System and from the arXiv e-print archive, in order to test three possible explanations 
for the citation advantage of papers that are freely available on the web. They found no causal link between open-access itself and citations. They did find a strong {\it early-access} effect, whereby papers pick up citations because they are available sooner; and also a strong {\it self-selection} bias effect, whereby authors tend to promote their most important and, thus, the most citable articles, by posting them on the arXiv.

Dietrich~\cite{Dietrich2008, Dietrich2008a} analyzed another kind of publicity for research articles, namely, placement in the top slots of the daily astrophysics (astro-ph) listing for e-prints published on the arXiv:astro-ph server. He found that e-prints appearing at or near the top of the astro-ph mailings receive significantly more citations than those further down the list. He identifies two causes for this citation boost: the {\it visibility} effect as more people see the top few papers in the list, and the {\it self-promotion effect} by authors who tend to promote their most important works by carefully timing their submission to the server in order to land in the top slots. (The visibility effect is essentially the publicity hypothesis of Phillips et al., while the self-promotion effect is essentially the self-selection bias effect of Kurtz et al.) Dietrich was able to separate the two effects, due to a serendipitous situation whereby some articles were accidentally placed in the top positions of listings. He concluded that increased visibility contributes to higher citation counts, but not as much as self-promotion by authors.

Ginsparg and Haque~\cite{Ginsparg} confirmed and extended Dietrich's results, by analyzing downloads and citations for 
listings of papers in astrophysics (astro-ph) and two large subcommunities of theoretical high energy physics (hep-th and hep-ph) of the arXiv. They reported a strong correlation between article position in these initial announcements and later citation counts, due primarily to intentional {\it self-promotion} by authors. Articles that fortuitously appeared near the top also received more citations than had they appeared in a lower position, due to a {\it visibility bias}. The increase in citations due to the visibility bias was smaller than that due to the self-promotion bias. By correlating download and citation data, the authors were also able to conclude that citations appear because of readership and lead to further readership.


Wainer, Eckmann, \&  Rocha \cite{Wainer} analyzed the citations of papers in prestigious Computer Science conferences that were selected on the basis of peer review as ``best'' for their year prior to presentation at the conference and publication. They found strong evidence that the selected papers are cited more than random papers at the conference, with a citation advantage that remained stable for at least two years following publication. They also reported that a significant number of the ``best'' papers are among the top cited papers in the conference.

Previously, we reported~\cite{PRB-Editorial} a citation advantage of papers highlighted in {\it Physical Review B} (PRB) as Editors' Suggestions. We also found that Editors' Suggestions were six times more likely than other PRB papers to be cited in the top 1\% in the field of physics, 
in annual lists of highly cited papers curated by Clarivate Analytics~\cite{Clarivate-HC}.

The potential of editorial selections for identifying outstanding research has been noted by experts at the European Research Council, Europe's prestigious funding agency of frontier research. Mugabushaka, Sadat \& Dantas Faria \cite{Mugabushaka_1} identified ``editorial highlights'' as one type of ``recognition channel'' and created an open dataset \cite{ERC_dataset} of editorial highlights for use in bibliometrics research and evaluative bibliometrics, including papers that received highlighting in {\it Science} as ``Breakthroughs of the year'' or {\it La Recherche} as ``les 10 découvertes de l’année'', as well as coverage in {\it Nature}  as ``research highlights'' and in {\it Science} as ``editor's choice''. 
In followup work, Sadat and Mugabushaka \cite{Mugabushaka_2} observe that editorial recommendations carry a precious expert judgement of scientific quality that is missing from the current practices of assessing research impact shortly after publication, like altmetrics. They proposed comparing these editorial recommendations with currently established measures of research impact. As a case in point, they report that 5\% of the recommended articles of Nature’s ``Research Highlights'' were associated with projects funded by the European Research Council.

%
%
\begin{table}[!t]
\caption{Highlighting platforms, by publisher, type, and write-up coverage}
\label{tab:1}       
%
%
\begin{tabular}{p{3cm}p{3cm}p{3cm}p{3cm}}
\hline\noalign{\smallskip}
Publisher & Highlight name & Type (intra- or inter-) & Write-up type  \\
\noalign{\smallskip}\svhline\noalign{\smallskip}
AAAS$^a$ & Editors' Choice  & inter-highlight & summary \\
AAAS & Perspective  & both & expert commentary\\
NPG$^b$ & Research Highlight & inter-highlight & summary\\
NPG & News \& Views & both & summary\\
APS$^c$ & Editors' Suggestion  & intra-highlight & none\\
APS & Viewpoint  & intra-highlight & expert commentary\\
APS & Synopsis  & intra-highlight & summary\\
IOP$^d$ & IOPselect  & intra-highlight & none\\
OSA$^e$ & Spotlight on Optics  & intra-highlight & none\\
ACS$^f$ & ACS Editors' Choice  & intra-highlight & none\\
NAS$^g$ & Journal club  & intra-highlight & summary \\
NAS & Commentary  & intra-highlight & expert commentary\\
JPS$^h$ & Editors' Choice  & intra-highlight & none\\
\noalign{\smallskip}\hline\noalign{\smallskip}
\end{tabular}
$^a$ American Association for the Advancement of Science \\
$^b$ Nature Publishing Group (Springer Nature)\\
$^c$ American Physical Society\\
$^d$ Institute of Physics\\
$^e$ Optical Society of America\\
$^f$ American Chemical Society\\
$^g$ National Academy of Sciences\\
$^h$ Japanese Physical Society\\
\end{table}
%




\section{Methods}
\label{sec:Methods}

We study papers published in {\it Physical Review Letters} (PRL), the flagship journal of the American Physical Society (APS). 
The highlighting platforms include intra-highlights in APS venues (PRL itself and the APS publication {\it Physics}~\cite{APS-physics}) and inter-highlights in other journals and publishers, such as {\it Nature Physics} and {\it Nature}. The intra-highlights are (i) journal cover in PRL (i.e., choice of paper in the weekly cover of the journal), (ii) {\it Editors' Suggestion} in PRL, (iii) {\it Focus} in {\it Physics}, (iv) {\it Synopsis} in {\it Physics}, and (v) {\it Viewpoint} in {\it Physics}. The inter-highlights are (i) {\it Research Highlight} and (ii) {\it News \& Views}.

To assess the citation performance of  highlighted papers, we use citation data from the Web of Science of Clarivate Analytics. We analyze citation data in three ways. We perform a multiple linear regression for the various highlighting platforms, using citation data from 1--10 years after publication, and compare regression coefficients. We also display box plots for the citation distributions of each highlighting platform, and compare medians. Finally, we compare the overlap of papers in each  platform with the top 1\% of cited papers, classified as highly cited papers in the Essential Science Indicators of Clarivate Analytics~\cite{Clarivate-HC}.

\subsection{Examined highlighting platforms}
\label{subsec:2.1}

Since every highlighting platform is different, it is useful to briefly describe the platforms we analyzed in this study. 

\begin{itemize}

\item{  {\it PRL cover image}: The journal cover is chosen by the editors for aesthetic reasons mainly (beautiful figures in the paper). Placement of a paper in the cover does not generally reflect selection on the basis of importance.}

\item{ {\it Editors' Suggestions}. First launched in 2007, these are generally chosen for ``potential interest in the results presented and, importantly, on the success of the paper in communicating its message''~\cite{PRL-Suggestions-2007}. While importance is not the only reason why a paper is selected, Editors' Suggestions fare well ``among many measures of importance, [...] impact, and intrinsic interest''~\cite{PRL-Suggestions-2017}.}



\item{ {\it Viewpoints}: First launched in 2008, these are commentaries commissioned by the {\it Physics} editors and written by experts. They explain why a paper is important to the field~\cite{APS-physics}.}

\item{ {\it Focus stories}: These are journalist-written news stories aiming to explain some of the latest research to the broadest possible audience~\cite{APS-physics}, i.e., to the educated non-physicists. First launched in 1998, Focus merged with {\it Physics} in 2011, where it continues to produce the same style of articles as before.}

\item{ {\it Synopses}: First launched in 2008, these are short summaries of newsworthy results written by journalists and {\it Physics} staff~\cite{APS-physics}.} 

\item{ {\it Research Highlights}: These are short summaries of papers written by journal editors in the {\it Nature} journals.} 

\item{ {\it News \& Views}: These are commentaries by experts or (less often) by journal editors in the {\it Nature} journals.}

Note that Viewpoints, Synopses, and Focus stories are mutually exclusive. A paper highlighted by a Viewpoint does not get considered also for Synopsis or Focus.

\end{itemize}

\subsection{What is importance?}
\label{subsec:2.2}

Since a central aspect of this work is the citation analysis of manuscripts marked ``important'' by the editors, it is necessary to say a few words about what we mean by importance and how such assessments are made in practice. We could use the working definition by the late Jack Sandweiss, former Lead Editor and Chairman of the Editorial Board of Phys. Rev. Letters, who used to say that ``an important result is one that researchers in the field should not miss, while those in related fields would be interested in''~\cite{Jack_Sandweiss}. A more detailed description is given in the journal's website: ``Important results are those that substantially advance a field, open a significant new area of research, or solve---or take a crucial step toward solving---a critical outstanding problem and thus facilitate notable progress in an existing field.''~\cite{PRL_policies}. 
Of course, an element of subjectivity is always present in any notion of importance, and a judgment call by the editor is implied in the words ``substantially,'' ``significant,'' and ``crucial'' in the above definition of importance.  
Nevertheless, the above descriptions of importance are {\it practically} useful to editors in at least {\it shortlisting} potentially important papers, because they help them develop the mindset and set the stage for what kind of papers to look for. 

Once a paper is accepted in Phys. Rev. Letters, the editors may shortlist it for potential highlighting. Decisions for Editors' Suggestion are made at the journal level. For further highlighting in {\it Physics}, the papers nominated by journal editors are discussed in a committee of {\it Physics} editors, who decide which papers get a Synopsis and which papers warrant further consideration as Viewpoints, in which case external experts are consulted, both for their opinion on the paper and on their availability as Viewpoint authors. So, by the time a PRL paper has received a Viewpoint, it has normally gone through three additional layers of post-acceptance scrutiny: consideration as an Editors' Suggestion by the journal, consideration as a Viewpoint by the {\it Physics} editors, and final vetting by external experts.

\subsection{Variation in coverage}
\label{subsec:2.3}

\begin{figure}[b]
\sidecaption
\includegraphics[scale=.55]{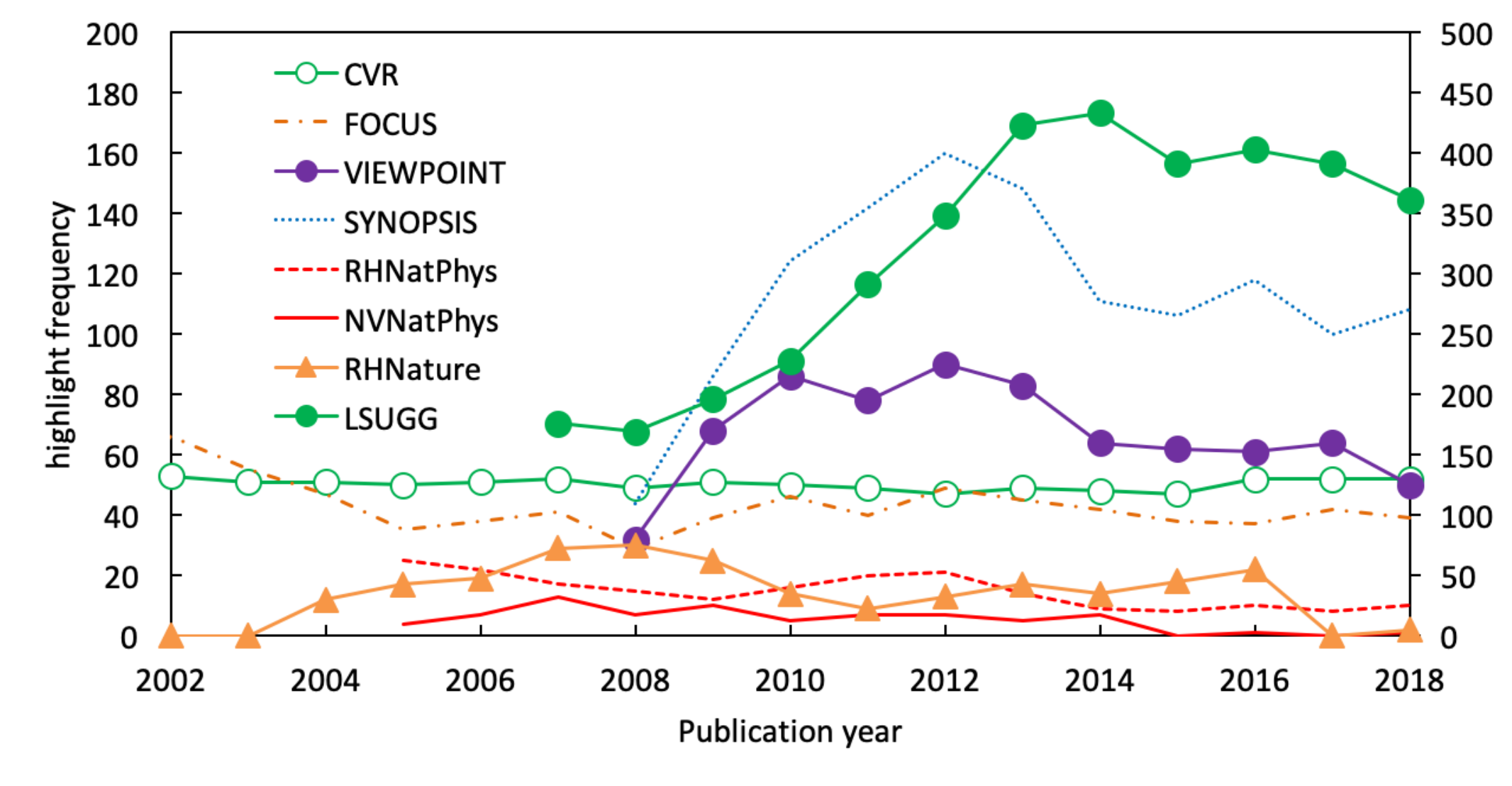}
%
%
\caption{Annual frequency of each highlighting platform. Editors' Suggestions (LSUGG) are plotted on the secondary (right) axis. Note: Editors' Suggestions started in 2007, Viewpoints and Synopses in 2008.}
\label{fig:time_series_1}       
\end{figure}

\begin{figure}[b]
\sidecaption
\includegraphics[scale=.55]{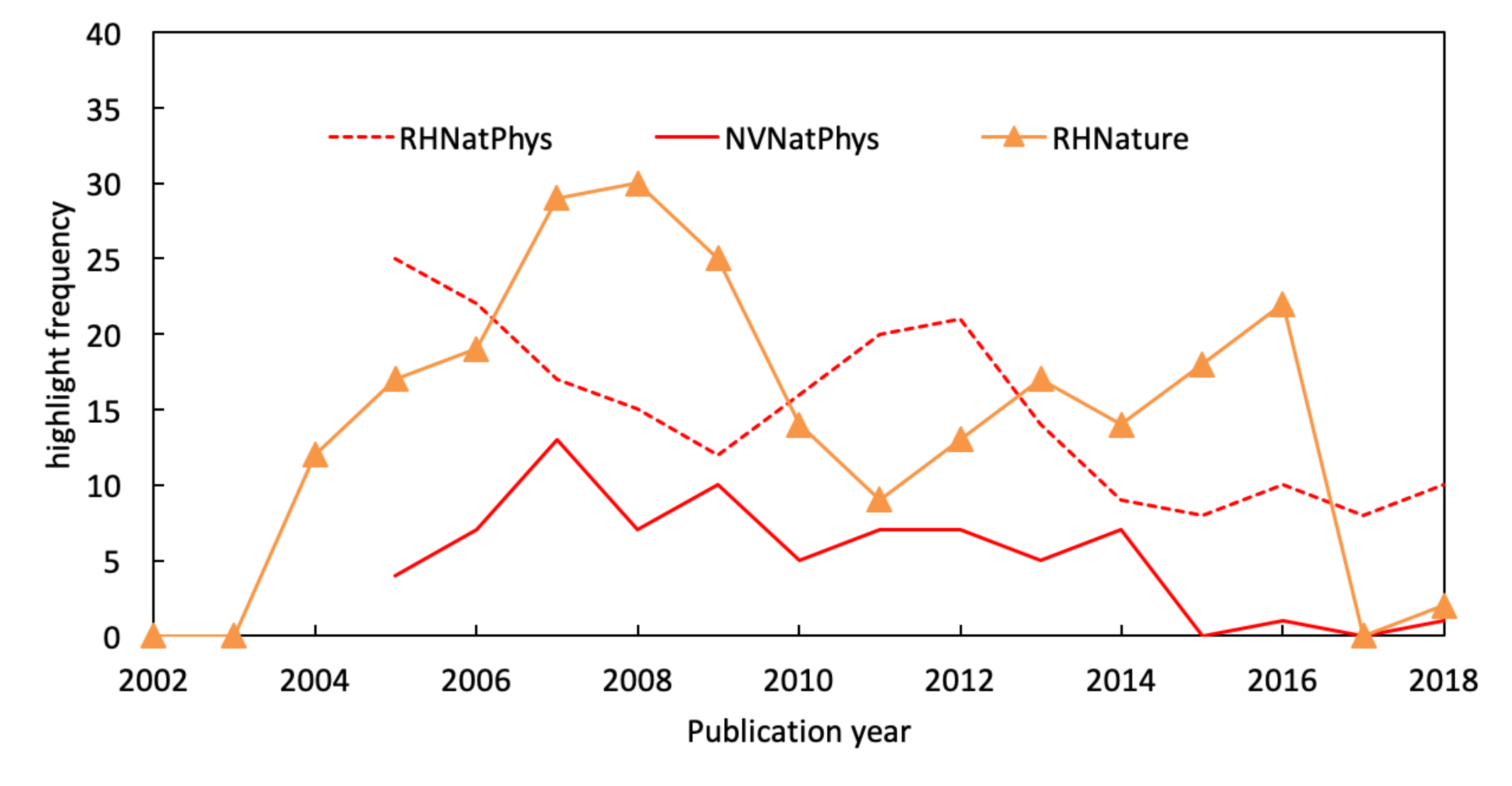}
%
%
\caption{Detail of Fig.\ref{fig:time_series_1}. Annual frequency of the inter-highlighting platforms analyzed here. 
}
\label{fig:time_series_2}       
\end{figure}

%
%
\begin{table}[!t]
\caption{Median annual frequency of each highlighting platform for PRL papers published from 2010--2018.}
\label{tab:2}       
%
%
\begin{tabular}{p{1cm}p{1cm}p{1.5cm}p{1.5cm}p{1.5cm}p{1.5cm}p{1.5cm}p{1.5cm}}
\hline\noalign{\smallskip}
CVR & Focus & Viewpoint & LSUGG & Synopsis & RHNatPhys & NVNatPhys & RHNature \\
\noalign{\smallskip}\svhline\noalign{\smallskip}
50 & 40 & 64 & 354 & 111 & 14 & 6 & 14 \\
\noalign{\smallskip}\hline\noalign{\smallskip}
\end{tabular}
\end{table}
%

%
%
%

In Fig.~\ref{fig:time_series_1} we show the annual frequency of each highlighting platform used in this study. Editors' Suggestions are most frequent (about 30 per month), followed by Synopses (10 per month) and Viewpoints (5 per month). On the other hand, inter-highlights are the least frequent platforms, with an annual median count of just over a dozen (Research Highlights in Nature and Nature Physics) or a half-dozen (News \& Views in Nature Physics). See Table~\ref{tab:2}. Inter-highlights display also higher variation from year to year, reflecting shifting priorities or focus of {\it Nature} journal editors in covering the most newsworthy developments in physics, the entry of new journals in the physics market, etc. See Fig.~\ref{fig:time_series_2}.

\section{Results and Discussion}
\label{sec:2}

We analyze the citations of highlighted papers using three tools: (i) multiple linear regression, (ii)  boxplots of citation distributions, and (iii) lists of highly cited papers by Clarivate Analytics. Multiple linear regression is our central tool.  

\subsection{Multiple Linear Regression}
\label{subsec:1}

We perform Multiple Linear Regression on the citation data of the papers highlighted in the seven different highlighting platforms described in Sect.~\ref{subsec:2.1}. We collect citation data for each paper each year for a range (citation window) of 1--10 years after publication, in order to allow dynamic observation of the citation performance of highlights. Obviously, for recently published papers ($<10$ years), not all citation windows can be counted. 

\begin{table}[!t]
\caption{Multiple linear regression coefficients for the citation data of papers highlighted in the various platforms, over a range (cumulative) of 1--7 years after publication. The papers were published from 2008--2012. Shown in the 2nd column are the numbers of papers in each platform. CVR = PRL journal cover, LSUGG = PRL {\it Editors' Suggestion}, RHNatPhys = Research Highlight in {\it Nature Physics}, NVNatPhys = News \& Views in {\it Nature Physics}, RHNature = Research Highlight in {\it Nature}. Statistical significance: $^{*}$ denotes $P<0.05$, $^{**}$ denotes $P<0.01$, $^{***}$ denotes $P<0.001$. No star denotes $P>0.05$. 
} 
\label{tab:3}       
%
%
\begin{tabular}{p{2cm}p{1.5cm}p{1cm}p{1cm}p{1cm}p{1cm}p{1cm}p{1cm}p{1cm}}
\hline\noalign{\smallskip}
Highlight & \# papers & 1Y & 2Y & 3Y & 4Y & 5Y & 6Y & 7Y\\
\noalign{\smallskip}\svhline\noalign{\smallskip}
CVR	& 246 & 0.6	& 2.2 & 5.1$^{*}$ & 6.3$^{*}$ & 7.5$^{*}$ & 8.5 & 9.7 \\
FOCUS & 203 & -0.1 & 0.2 & 0.6 & 1.5 & 2.4 & 3.2 & 4.1 \\ 
VIEWPOINT & 354 & 10.3$^{***}$ & 20.4$^{***}$ & 29.0$^{***}$ & 37.4$^{***}$ & 44.3$^{***}$ & 51.3$^{***}$ & 58.0$^{***}$ \\
LSUGG & 1232 & 3.4$^{***}$ & 6.7$^{***}$ & 10.0$^{***}$ & 12.9$^{***}$ & 15.7$^{***}$ & 18.5$^{***}$ & 20.9$^{***}$ \\
SYNOPSIS & 556 & 1.7$^{**}$ & 4.0$^{***}$ & 5.5$^{***}$ & 7.2$^{***}$ & 8.6$^{***}$ & 9.8$^{**}$ & 10.7$^{*}$ \\
RHNatPhys & 84 & 2.4$^{*}$ & 5.8$^{*}$ & 8.4$^{*}$ & 11.4$^{*}$ & 15.1$^{*}$ & 19.0$^{*}$ & 23.8$^{*}$ \\
NVNatPhys & 36 & 2.9 & 6.3 & 8.7 & 10.7 & 13.0 & 14.7 & 15.8 \\
RHNature & 91 & 2.3$^{*}$ & 6.6$^{**}$ & 11.7$^{***}$ & 17.6$^{***}$ & 24.8$^{***}$ & 32.5$^{***}$ &	39.0$^{***}$ \\
\noalign{\smallskip}\hline\noalign{\smallskip}
\end{tabular}
\end{table}
%

%
\begin{figure}[b]
\sidecaption
\includegraphics[scale=.55]{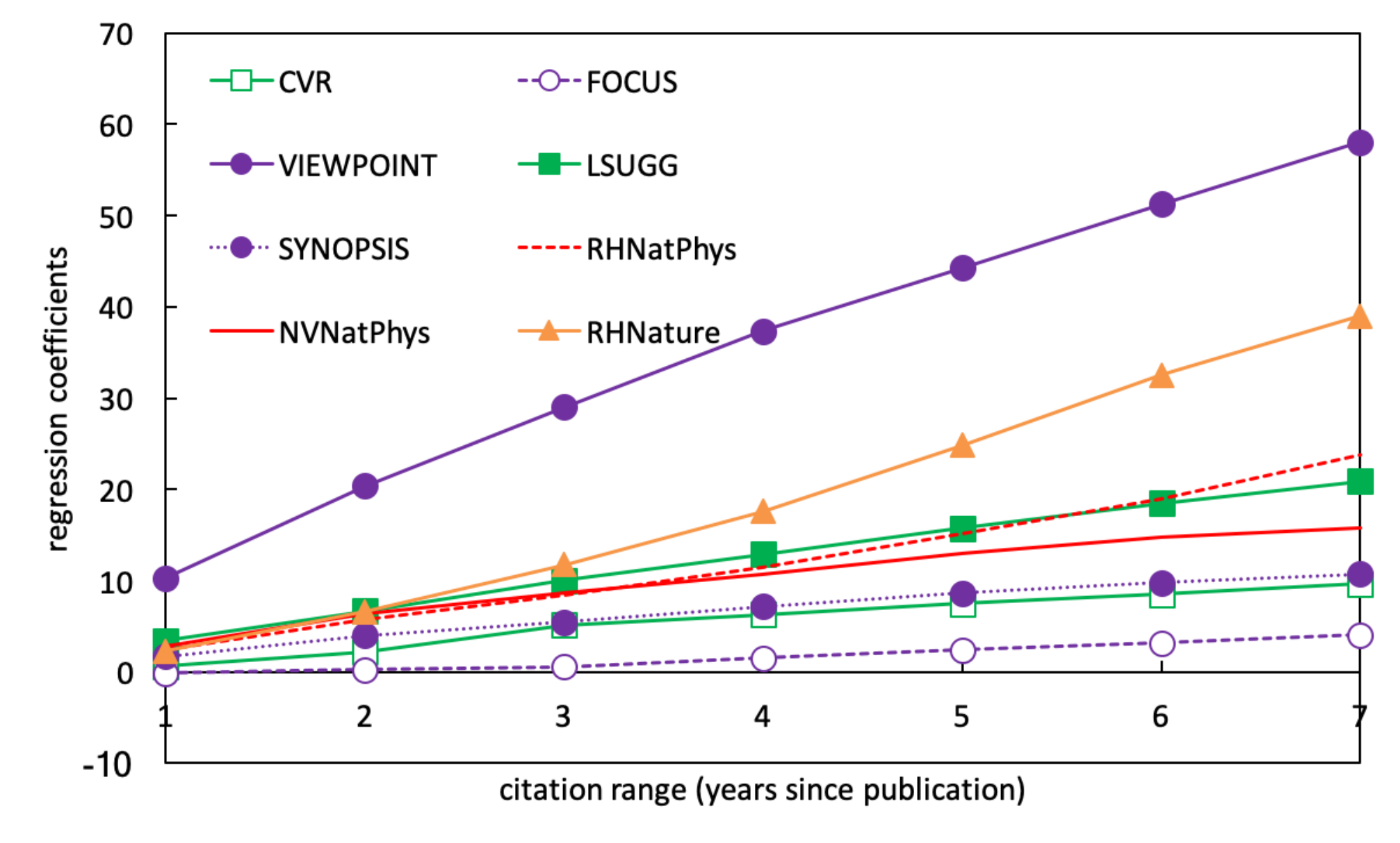}
%
%
\caption{Coefficients of multiple linear regression for each highlighting platform. Papers were published from 2008--2012. Citations were collected from 1--7 years after publication.}
\label{fig:coeffs_2008-2012}       
\end{figure}

As a first sample to analyze, we chose papers published from 2008--2012, and counted their cumulative citations from 1--7 years after publication. This amounts to 246 papers in the journal cover, 203 in Focus, 354 Viewpoints, 1232 Editors' Suggestions, 556 Synopses, 84 Research Highlights in {\it Nature Physics}, 36 News \& Views in {\it Nature Physics}, and 91 Research Highlights in {\it Nature}. Note that Viewpoints and Synopses first started in 2008, and Editors' Suggestions in 2007. In Table~\ref{tab:3} we display the regression coefficients for each highlighting type. We use a single star (*) for values with $P<0.05$, two stars (**) for $P<0.01$, and three stars (***) for $P<0.001$. We find statistical significance for all data of Viewpoints, Suggestions, Synopses, and Research Highlights in {\it Nature}, and for 3 years of Covers. The data are also plotted in Fig.~\ref{fig:coeffs_2008-2012} for  clarity.


The first thing to note from Table~\ref{tab:3} and  Fig.~\ref{fig:coeffs_2008-2012} is the strong association between a Viewpoint and high citations. In all citation windows, i.e., from 1 to 7 years after publication, the Viewpoint marker is clearly a stronger predictor of citations than any other marker of publicity. All else being equal, a Viewpoint marker predicts 10 additional citations to a PRL paper during the first year after publication, 44 citations within 5 years, and  58 citations within 7 years. This strong position of Viewpoints as the leading marker for citation accrual is found across all years and samples, so it is a very stable result. 


The second strongest association with citations is found for Research Highlights in {\it Nature}. 5 years within publication, this marker predicts almost 25 additional citations, all else being the same. In the first 1--3 years since publication, this effect is not clearly pronounced compared to the other markers, but from the 4th year onward it can be clearly seen. 

The next strongest effect is found for Research Highlights in {\it Nature Physics} and Suggestions, whose behavior is similar, so we group them together; and after that, for Synopses and Covers, which we also group. 


Broadly speaking, we observe the following stratification of citation advantage for the different highlights:
\begin{equation}
\text{Viewpoint} 
> \text{RH {\it Nature}} >
\left\{
    \begin{array}{ll}
        \text{Suggestion} \\
        \text{RH {\it Nat. Physics}} \\ 
        \text{\; \; \; [N\&V {\it Nat. Physics}]} \\
    \end{array}
\right\}
> 
\left\{
    \begin{array}{ll}
        \text{Synopsis}  \\
        \text{Cover}  \\
        \text{\; \; \; [Focus]} 
    \end{array}
\right\}.\\
\label{eq:A01}
\end{equation}
Here, we have grouped together highlights in terms of decreasing citation advantage. In the first group are Viewpoints, which are cited the most. In the second group are Research Highlights in {\it Nature}. In the third group we have Suggestions and Research Highlights in {\it Nature Physics}. And in the fourth group we have Synopses and Covers. For completeness, we included News \& Views in {\it Nature Physics} and Focus papers, but put them in brackets, since their coefficients lack statistical significance.

Interestingly, the stratification in Eq.~\ref{eq:A01} follows also a hierarchical pattern of decreasing scrutiny, with regard to importance, during peer review. First, Viewpoints and Suggestions are intra-highlights, so the editors who select them have the full benefit of both internal and external to the journal review. Between these two, Viewpoints are clearly vetted more, since they are discussed not only among the PRL journal editors once the papers are accepted for publication, but also among a committee of Physics editors who then seek additional advice from external experts on whether such papers merit highlighting as Viewpoints. Synopses and Viewpoints are mutually exclusive, and since Viewpoints receive the highest level of scrutiny, it follows that Synopses do not meet the same standard in terms of importance during the vetting process---indeed, many intellectually curious, `fun' or elegant papers are selected for Synopsis. Focus stories are  
aimed at an audience of educated non-physicists, and papers are often chosen based on educational value and intrinsic interest to non-specialists rather than on scientific merit, so, again, importance is not the key criterion for selection. We find that a Focus marker has no significant prediction on a paper's citations. (We discuss Covers further below.)

Research Highlights in {\it Nature} or {\it Nature Physics}, and News \& Views in {\it Nature Physics} are all inter-highlights, which makes their selection more challenging to begin with, since the NPG editors do not have access to the peer review files from PRL to aid their decisions. These journals are also more constrained in terms of journal space since their highlights cover several journals, and, in the case of {\it Nature}, several fields. So it does not seem surprising that these highlighting platforms yield a smaller citation advantage than Viewpoints, for which it is reasonable to assume that papers receive the highest scrutiny among all markers presented here. As for the relative difference in citations between Research Highlights in {\it Nature} or {\it Nature Physics}, this may be due to the fact that {\it Nature} is more selective than {\it Nature Physics} since it covers all fields of science, and its highlighting slots may therefore be more vetted.  

A Cover marker predicts a statistically significant but small citation boost. All else being equal, placement in the journal cover predicts no more than 1.5 additional citations per year (7.5 citations within 5 years of publication, almost 10 citations within 7 years). The small effect of cover placement may come as a surprise to some. For instance, the author has first-hand experience, from institutional visits throughout the world, of the pride many authors take in seeing their work appear in the cover of PRL. Indeed, authors often interpret placement in the cover as an editorial endorsement of their paper's higher importance among other papers in a journal issue. However, selection of a cover figure for PRL, as for several other scientific journals, is primarily done for aesthetics and does not typically imply an editorial endorsement of higher scientific merit. (Of course, if the editors have qualms about a paper's results or conclusions they are less likely to select it for the cover.) While the reasons behind cover selection seem to be largely lost to the community (according to our anecdotal experience), the prediction of the cover marker on citations actually falls in line with the editors' opinion: papers in the cover are cited only marginally more than other papers, all other things being the same. So, this type of {\it  accidental or serendipitous} publicity brings a small citation advantage---in direct analogy to the effect of {\it visibility} bias reported by Dietrich ~\cite{Dietrich2008, Dietrich2008a} and Ginsparg and Haque~\cite{Ginsparg}, for papers that accidentally end up in the top slot of the daily arXiv email listings.However, the citation advantage is clearly greater when publicity is deliberate and results from an endorsement of the paper's merit formed through peer review, as in a Viewpoint or any marker presented here other than Focus. Viewed from this lens, we can understand the {\it self-promotion} bias in Dietrich ~\cite{Dietrich2008, Dietrich2008a} and Ginsparg and Haque~\cite{Ginsparg} for papers whose authors deliberately and carefully engineered their placement in the top slot
or arXiv lists, as a form of {\it internal} review: Clearly, authors aim to push their better papers in the top slot of arXiv listings, and this self-selection  reflects an {\it endorsement} of higher quality and results in a citation advantage.

So, publicity alone does help in terms of citations, but does not make a lot of difference unless it is supported by an endorsement, formed through peer review (internal and/or external), that the paper has above-average merit. 

To recap, we observe that increased editorial scrutiny of a journal's papers in terms of importance is associated with higher citation counts. Our findings are consistent with the notion that citations follow or mirror peer review~\cite{REF-rainy-Sunday} \cite{metrics-peer-review-Traag} \cite{PRB-Editorial}. Our analysis takes this premise a step further, since it shows that we can allocate relative degrees of importance and correspondingly predict citation accrual for papers published in the same journal but publicized by different highlighting platforms. 

\subsection{Box plots and medians}
\label{subsec:2}

In the previous subsection, we used multiple linear regression to study the association between the various highlighting platforms and the citations of the highlighted papers, in order to identify the different prediction of each marker. Another way to visualize our data is to compare the citation distributions of the highlighted sets of papers. One caveat here is that some papers are co-highlighted in more than one platform (say, a Suggestion, Viewpoint, and Cover) and are thus present in all the corresponding distributions. With this caveat in mind, we proceed.

%
%
\begin{figure}[b]
\sidecaption
\includegraphics[scale=.35]{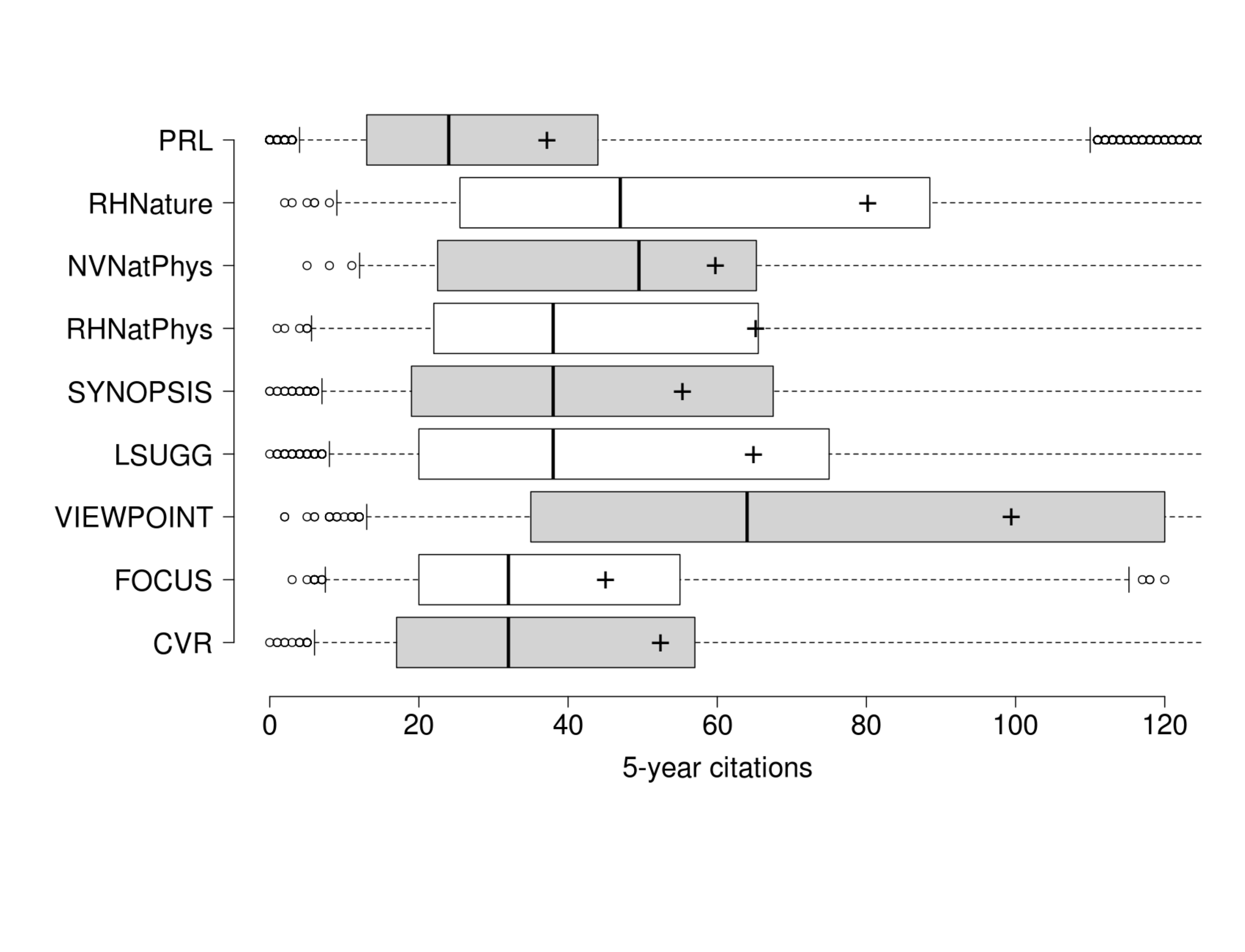}
%
%
\caption{Boxplots of citations of highlighted papers, plotted horizontally. Publication years = 2008--2014. Citations counted over a 5-year window since publication.  The horizontal extent of each box spans the {\it interquartile range} IQR = Q1--Q3 (i.e., from the 25$^{\rm th}$--75$^{\rm th}$ percentile). Medians are shown as line segments, means as crosses. Whiskers extend from the 5th to the 95h percentile (most 95th-percentile whiskers are not shown in this truncated plot). Outliers are shown as dots beyond the extent of whiskers. Alternate white/gray shading is used simply to increase legibility.}

\label{fig:boxplot_5Y_detail}       
\end{figure}
Because citation distributions are usually highly skewed, their comparison is done by visual inspection of the corresponding boxplots. A boxplot is a graphical depiction of a 5-number summary of the distribution, which in the Altman  convention~\cite{Altman} that we follow here includes the 5th percentile, the 25th percentile (or 1st quartile), the median, the 75th percentile (or third quartile), and the 95th percentile. Outlier data points that lie below the 5th and above the 95th percentiles are shown as dots. 

For example, in Fig.~\ref{fig:boxplot_5Y_detail} we show the boxplots for the 5-year citations of each highlighted group, for papers published from 2008-2014. Once again, Viewpoints are clearly cited the most, whether one compares medians, means (shown as crosses), or interquartile ranges (Q1--Q3). If we rank the remaining groups by medians, in order to avoid the distorting effect of outliers (highly cited papers) on the means~\cite{QSS}, we have News \& Views in {\it Nature Physics} and Research Highlights in {\it Nature} closely together, followed by Suggestions, Synopses and Research Highlights in {\it Nature Physics} that are also on par. Finally, we have Cover and Focus articles.

\begin{figure}[b]
\sidecaption
\includegraphics[scale=.55]{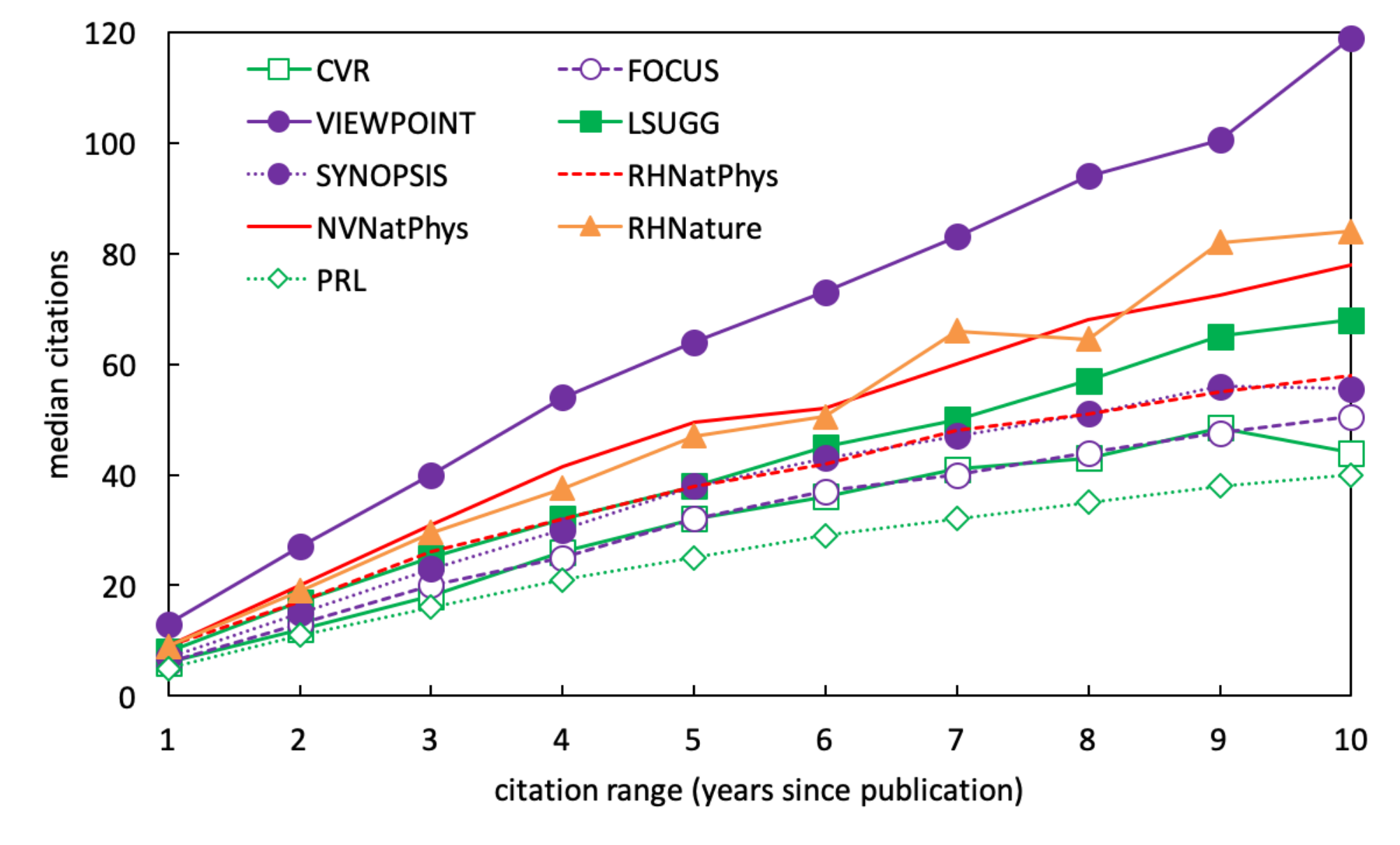}
%
%
\caption{Median citations for each highlighting platform and for the PRL journal. Papers were published from 2008--2018. Citations were collected from 1--10 years after publication.}
\label{fig:median_citations}       
\end{figure}

\begin{figure}[b]
\sidecaption
\includegraphics[scale=.55]{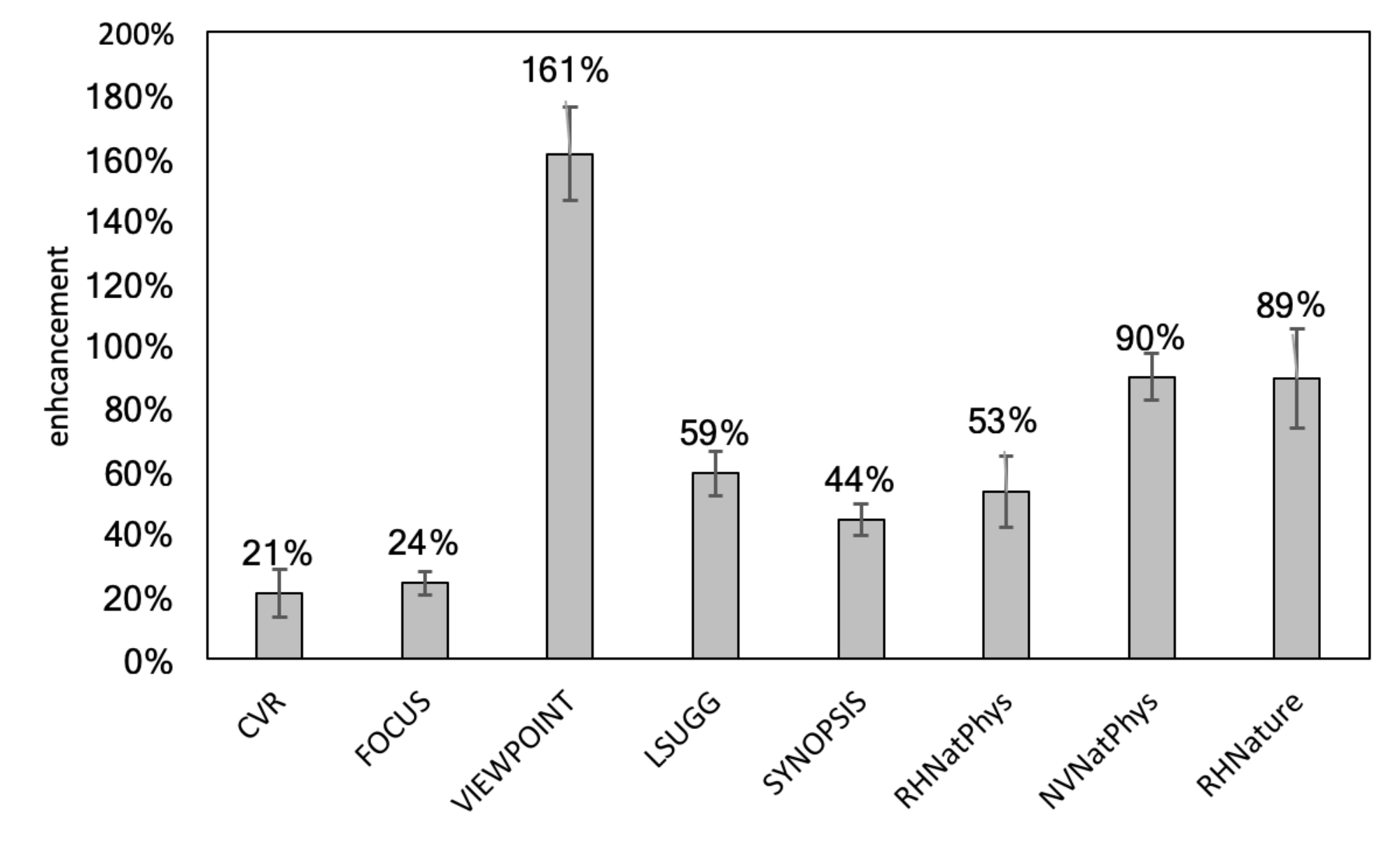}
%
%
\caption{Average value of enhancement of median citations for each highlighting platform compared to the PRL journal. Papers were published from 2008--2018. Citations were collected from 1--10 years after publication. Error bars show 99\% confidence intervals.}
\label{fig:median_enhancement}       
\end{figure}

We checked for statistical significance in the difference between any two medians among these 8 distributions of highlights using the nonparametric Mann-Whitney U test. (With the added caveat that Viewpoints, Synopses, and Focus articles are mutually exclusive and hence not quite independent.)
Our null hypothesis is that the medians are equal and the one-sided alternative is that one median is greater than the other, at 5\% significance level. For the 5-year citation distributions of Fig.~\ref{fig:boxplot_5Y_detail}, we could thus establish that, at 5\% significance level: \\
$\bullet$ the citation median for an article with a Viewpoint is greater than for all other articles; \\
$\bullet$ the citation median for articles in Research Highlights in {\it Nature} is greater than for Suggestions, Synopses, Covers, or Focus articles; and \\
$\bullet$ the citation median for articles in Research Highlights in {\it Nature Physics}, News \& Views in {\it Nature Physics}, Suggestions, and Synopses is greater than for Covers, or Focus articles.

The above ranking of citation medians essentially confirms the stratification of Eq.~\ref{eq:A01}, which is reassuring. Any minor adjustments (e.g., Synopsis faring better than Covers) reflect small, gradual shifts and are thus inconsequential.

Fig.~\ref{fig:boxplot_5Y_detail} utilizes just a part of our full dataset, i.e., papers published from 2008--2014 and cited over the next 5 years. To utilize the full dataset and reduce statistical uncertainties, we calculated the citation median for the citation range 1--10 years after publication for all combinations of publication years from 2008--2018. 
That is, to obtain the citation median for one year since publication, we took the median of citations from 2009, 2010, ..., 2019 to articles published respectively in 2008, 2009, ..., 2018, etc. For two years since publication, we took the median of citations from 2009--2010, ..., 2018--2019 to articles published respectively in 2008, ..., 2017, etc. 
See Fig.~\ref{fig:median_citations}. 

We then processed these data a little further, as follows. For each year in the citation range and each highlighting marker, we calculated the relative increase of the citation median compared to the journal median, and took the average over all publication years. This is how we calculated the average enhancement shown in Fig.~\ref{fig:median_enhancement}. For example, papers in the journal cover have a citation median that is on average 21\% higher than the baseline (the whole journal) over the 10-year citation range.

Put together, the results from the figures presented here broadly confirm our findings from the previous subsection. Intra-highlights for importance (Viewpoints) bring the most citations (1st group), followed by inter-highlights for importance (News \& Views in {\it Nature Physics}, Research Highlights in {\it Nature} and {\it Nature Physics}) and Editors' Suggestions (2nd group), followed by accidental publicity papers (covers) and Focus stories (3rd group). 

Even though box plots do not correct for 
co-highlighting, as we mentioned before, the stratification of citations for Viewpoints, Suggestions, Synopses, and Focus, as well as Research Highlights in {\it Nature} and {\it Nature Physics}, broadly agrees with the regression analysis. This makes sense, for the reasons discussed in the previous subsection. 

It is worth noting that the Suggestions collection contains many highly cited papers, even though its median (and mean) are considerably lower than for Viewpoints.

\subsection{Inclusion in Clarivate list of highly cited papers}
\label{subsec:3}

Here, we ask the question, {\it Can we use highlighting markers as predictors of a paper being highly cited?} This question takes us beyond the analysis of multiple linear regression, where we looked at how many citations the `average' paper picked upon being highlighted. Placement of a paper in a {\it highly} cited list is identifying potential for extreme, not average, citation performance. 

As a benchmark for highly cited papers, we use the {\it Highly Cited Papers (HCP)}~\cite{Clarivate-HC} indicator of Clarivate Analytics. These papers are the top 1\% cited in their subject per year. Lists of HCP papers going back 10 years are available for download from the Essential Science Indicators of Clarivate Analytics. We downloaded the list 
of HCP papers in physics published from 2010--2018, among which there are 1371 papers published in PRL. For each highlighting marker we calculate 
how often it ``predicts'' placement in the HCP list, i.e., how many highlighted papers are highly cited; this is the positive predictive value, or {\it precision}. We also calculate how often the HCP list contains marked papers, i.e., how many highly cited papers are highlighted; this is the true positive rate or {\it sensitivity}. The geometric mean of  these two quantities (precision and sensitivity) is the {\it F1 score}~\cite{F1}, which is a measure of a marker's accuracy. See Table~\ref{tab:5}.

Evidently, 29\% of Viewpoint papers are placed in the highly-cited-papers list of Clarivate Analytics. So, Viewpoints have the highest precision, followed by Research Highlights in {\it Nature} (23.9\%), Research Highlights in {\it Nature Physics} (16.4\%), and Editors' Suggestions (14.3\%). The hierarchical pattern of Eq.~\ref{eq:A01} in terms of decreasing scrutiny is thus reproduced. Again, we find that highlighting for importance correlates with citations even for the extreme case of top-1\% cited papers, as in the HCP lists. 

With regard to sensitivity, the inter-highlighting platforms in {\it Nature} and {\it Nature Physics} score much lower than the intra-highlights, as expected (see below). For example, just under 0.3\% of HCP papers in PRL are highlighted by a News \& Views item in {\it Nature Physics}. For comparison, 3.2\% of the PRL's HCP papers are PRL Covers, and 34.1\% of PRL's HCP papers were marked by a Viewpoint. Why is this expected? Because inter-highlighting platforms such as Research Highlights and News \& Views have many journals to cover and a limited number of slots. So, {\it Nature} or {\it Nature Physics} can only highlight a few PRL papers at any time. At the same time, the high sensitivity values of Viewpoints and especially Editors' Suggestions are remarkable: More than 1 in every 3 HCP papers in PRL is an Editors' Suggestion. This result echoes our earlier remark that the Suggestions list contains many highly cited papers. 

From the precision and sensitivity values we obtain the F1 score (it is their geometric mean). The F1 score can be viewed as the overall accuracy of a highlighting marker's predictive power for placement in the HCP list. As shown in Table~\ref{tab:5}, Editors' Suggestions, Viewpoints, and Synopses dominate this score, with values 0.2, 0.18, and 0.1, respectively.

%
%
\begin{table}[!t]
\caption{Summary statistics for precision (or positive predictive value), sensitivity (or true positive rate), and F1 score, for the performance of each highlighting platform with regard to identifying (predicting) highly cited papers. Publication years, 2010--2018. 1371 papers in PRL were listed as highly-cited-papers (HCP) by Clarivate Analytics.}
\label{tab:5}       
%
%
\begin{tabular}{p{2cm}p{1cm}p{3.5cm}p{3cm}p{2cm}}
\hline\noalign{\smallskip}
Highlight & Count & Precision & Sensitivity & F1 score \\
 &  & (positive predictive value) & (true positive rate) &  \\
\noalign{\smallskip}\svhline\noalign{\smallskip}
CVR	& 446 & 0.099 & 0.032 & 0.048 \\
FOCUS & 378 & 0.063 & 0.018 & 0.027 \\
VIEWPOINT & 638 & 0.290 & 0.135 & 0.184 \\
LSUGG & 3269 & 0.143 & 0.341 & 0.201 \\
SYNOPSIS & 1117 & 0.117 & 0.096 & 0.105 \\
RHNatPhys & 116 & 0.164 & 0.014 & 0.026 \\
NVNatPhys & 33 & 0.121 & 0.003 & 0.006 \\
RHNature & 109 & 0.239 & 0.019 & 0.035 \\
\noalign{\smallskip}\hline\noalign{\smallskip}
\end{tabular}
\end{table}

\section{Conclusions}
\label{sec:conclusions}

We have presented an analysis on the citation advantage of papers from {\it Physical Review Letters} (PRL) that are highlighted in various platforms, both within and beyond the journal. Among the various highlighting platforms we analyzed, we found that Viewpoints in {\it Physics} are the strongest predictors of citation accrual, followed by Research Highlights in {\it Nature}, Editors' Suggestions in PRL, and Research Highlights in {\it Nature Physics}. 

When a PRL paper is highlighted by a Viewpoint in {\it Physics} or a Research Highlight in {\it Nature}, its chances of being listed among the top 1\% cited papers in physics are about 1 in 4.

Our key conclusion is twofold. First,  highlighting for importance identifies a citation advantage. 
When editors select papers to highlight with importance as their main criterion, they can identify papers that end up well cited. Accidental or serendipitous publicity (i.e., mere visibility), whereby a paper is publicized for other reasons than importance, e.g., in the journal cover, gives a clearly smaller citation advantage.

Second, the stratification of citations for highlighted papers follows the degree of vetting of papers for importance during peer review (internal and external to the journal). So, citation metrics mirror peer review even {\it within} a journal. This implies that we can view the various highlighting platforms (e.g., Viewpoints) as predictors of citation accrual, with varying degrees of strength that mirror each platform's vetting level.


%
\begin{acknowledgement}
I am grateful to Hugues Chat{\'e}, Jerry I. Dadap, Paul Ginsparg, and Jessica Thomas for stimulating discussions, and to Richard Osgood Jr. and Irving P. Herman for hospitality. This work uses data, accessed through Columbia University, from the Web of Science and InCites Essential Science Indicators of Clarivate Analytics.

\bigskip{}
\noindent
{\bf Competing interests}

\noindent 
The author is an Associate Editor in {\it Physical Review B}, a Contributing Editor in {\it Physical Review Research}, and a Bibliostatistics Analyst at the American Physical Society. He is also an Editorial Board member of the Metrics Toolkit, a volunteer position. He was formerly an Associate Editor in {\it Physical Review Letters} and {\it Physical Review X}. The manuscript expresses the views of the author and not of any journals, societies, or institutions where he may serve.

\end{acknowledgement}

\input{references}


%% file: author1/references.tex
%
%
%
%

%% file: editor/appendix.tex
%
%
%

\chapter{Chapter Heading}
\label{introA} 

Use the template \emph{appendix.tex} together with the document class SVMono (monograph-type books) or SVMult (edited books) to style appendix of your book.

\section{Section Heading}
\label{sec:A1}
Instead of simply listing headings of different levels we recommend to let every heading be followed by at least a short passage of text. Further on please use the \LaTeX\ automatism for all your cross-references and citations.

\subsection{Subsection Heading}
\label{sec:A2}
Instead of simply listing headings of different levels we recommend to let every heading be followed by at least a short passage of text. Further on please use the \LaTeX\ automatism for all your cross-references and citations as has already been described in Sect.~\ref{sec:A1}.

For multiline equations we recommend to use the \verb|eqnarray| environment.
\begin{eqnarray}
\vec{a}\times\vec{b}=\vec{c} \nonumber\\
\vec{a}\times\vec{b}=\vec{c}
\label{eq:A01}
\end{eqnarray}

\subsubsection{Subsubsection Heading}
Instead of simply listing headings of different levels we recommend to let every heading be followed by at least a short passage of text. Further on please use the \LaTeX\ automatism for all your cross-references and citations as has already been described in Sect.~\ref{sec:A2}.

Please note that the first line of text that follows a heading is not indented, whereas the first lines of all subsequent paragraphs are.

%
\begin{figure}[t]
\sidecaption[t]
\includegraphics[scale=.65]{figure}
%
%
\caption{Please write your figure caption here}
\label{fig:A1}       
\end{figure}

%
\begin{table}
\caption{Please write your table caption here}
\label{tab:A1}       
%
%
\begin{tabular}{p{2cm}p{2.4cm}p{2cm}p{4.9cm}}
\hline\noalign{\smallskip}
Classes & Subclass & Length & Action Mechanism  \\
\noalign{\smallskip}\hline\noalign{\smallskip}
Translation & mRNA$^a$  & 22 (19--25) & Translation repression, mRNA cleavage\\
Translation & mRNA cleavage & 21 & mRNA cleavage\\
Translation & mRNA  & 21--22 & mRNA cleavage\\
Translation & mRNA  & 24--26 & Histone and DNA Modification\\
\noalign{\smallskip}\hline\noalign{\smallskip}
\end{tabular}
$^a$ Table foot note (with superscript)
\end{table}
%

%% file: editor/glossary.tex
%
%

\Extrachap{Glossary}

Use the template \emph{glossary.tex} together with the Springer Nature document class SVMono (monograph-type books) or SVMult (edited books) to style your glossary\index{glossary} in the Springer Nature layout.

\runinhead{glossary term} Write here the description of the glossary term. Write here the description of the glossary term. Write here the description of the glossary term.

\runinhead{glossary term} Write here the description of the glossary term. Write here the description of the glossary term. Write here the description of the glossary term.

\runinhead{glossary term} Write here the description of the glossary term. Write here the description of the glossary term. Write here the description of the glossary term.

\runinhead{glossary term} Write here the description of the glossary term. Write here the description of the glossary term. Write here the description of the glossary term.

\runinhead{glossary term} Write here the description of the glossary term. Write here the description of the glossary term. Write here the description of the glossary term.